\documentclass[letters,useAMS,usenatbib,usegraphix]{mnras}
\usepackage{amsmath}
\usepackage{times}
\usepackage{aas_macros}
\usepackage{graphicx}
\usepackage[normalem]{ulem}
\usepackage[usenames]{color}
\usepackage{amsmath}
\usepackage{amssymb}

\newcommand\msun{M_{\odot}}
\newcommand\Zsun{Z_{\odot}}
\newcommand\zform{z_\mathrm{f}}
\newcommand\tform{t_\mathrm{f}}
\newcommand\zm{z_\mathrm{m}}
\newcommand\tm{t_\mathrm{m}}
\newcommand\Zc{Z_\mathrm{p}}
\newcommand\mgal{M_\mathrm{gal}}

\title[BBH merger host galaxies]{When and where did GW150914 form?}
\author[A. Lamberts et al.]{A. Lamberts$^{1}$\thanks{E-mail:
lamberts@caltech.edu}, S. Garrison-Kimmel$^{1}$, D. R. Clausen$^{1,2}$ and P. F. Hopkins$^{1,2}$\\
$^{1}$TAPIR, MC 350-17, California Institute of Technology, Pasadena, CA 91125, USA\\
$^{2}$Walter Burke Institute for Theoretical Physics, MC 452-48, California Institute of Technology, Pasadena, CA 91125, USA}

\begin{document}
\label{firstpage}
\date{\today}

\pagerange{\pageref{firstpage}--\pageref{lastpage}} \pubyear{2016}

\maketitle

\begin{abstract}

The recent LIGO detection of gravitational waves (GW150914), likely originating from
the merger of two $\sim 30 \msun$ black holes suggests progenitor stars of low metallicity ($[Z/\Zsun] \lesssim 0.3$), constraining when and where the progenitor of GW150914
may have formed.  We combine estimates of galaxy properties (star forming gas metallicity, star formation rate and merger rate)
across cosmic time to predict the low redshift black hole - black hole merger rate as a function of present day  host galaxy mass, $\mgal$, the
formation redshift of the progenitor system $\zform$ and different progenitor metallicities $\Zc$. For $\Zc\geqslant 0.1 \Zsun$, the signal is
dominated by binaries in massive galaxies with $\zform\simeq 2$ while below $\Zc\leqslant 0.1\Zsun$ most mergers come from binaries formed around $\zform\simeq 0.5$ in dwarf galaxies. Additional gravitational wave detections from merging massive black holes will provide constraints on the mass-metallicity relation and massive star formation at high redshifts.
\end{abstract}


\section{Introduction}

On 2015 September 14, both detectors from the Laser Interferometer Gravitational-Wave Observatory (LIGO) made the first direct detection of gravitational waves \citep{LIGO:2016_main}. The gravitational waves were emitted by two merging black holes of $M_1=36_{-4}^{+5}\msun$ and $M_2=29_{-4}^{+4}\msun$ located at redshift $z=0.09_{-0.04}^{+0.03}$. While the detection of black holes much heavier than any mass measured in X-ray binaries revives the study of the evolution of massive stellar binaries \citep{Mandel:2016,Marchant:2016}, the determination of host galaxies has been mostly ignored. So far, binary population syntheses (BPS) models have argued that GW150914 can only have formed in a low metallicity environment, below $0.25\Zsun$, most probably around $0.1\Zsun$ \citep{LIGO:2016_implications,Belczynski:2016}.

This strong limit on the progenitor metallicity allows one to determine in what type of galaxy and at what time the progenitors of massive ($M_1+M_2 \geqslant 40 \msun$, $M_1,M_2\geqslant 15\msun$) binary black holes (BBH) are born. While previous work has determined merger rates as a function of redshift \citep{Dominik:2013,Dvorkin:2016}, this work presents the first determination of the formation conditions for the massive BBH mergers we currently observe.  As the delay time between progenitor formation and BBH merger often exceeds several Gyr, one has to consider star formation through cosmic history to correctly model the progenitor population.

Low metallicity gas is typically found in high redshift galaxies or in local dwarf galaxies.  Using a two component model for the star formation and metallicity as a function of redshift, \citet{OShaughnessy:2010} showed that elliptical galaxies dominate BBH mergers hosts. Based on the redshift evolution of a mass-independent metallicity distribution with significant scatter, \citet{Belczynski:2016} suggests two roughly equally probable formation times for GW150914 around $z\simeq 3$ and below $z\simeq 0.2$. In this work  we use a complete,  redshift dependent mass-metallicity relation (MZR) consistent with recent high-redshift observations \citep{Erb:2006,Mannucci:2009}. Additionally, we explicitly account for galaxy mergers that bring low metallicity stars/black holes formed in low mass galaxies to higher mass galaxies, where the BBH mergers take place.

In order to determine the environment in which GW150914 formed, we assume the progenitors have the metallicity of the gas in which they form.  First, we determine the amount of low metallicity star formation through cosmic history (\S\ref{sec:lowZ}). Using a binary population synthesis (BPS) model, we then determine the delay time distribution for various progenitor metallicities (\S\ref{sec:mergers}). We finally combine both computations to determine where GW150914 most likely formed (\S\ref{sec:combine}) and discuss the implications for future detections (\S\ref{sec:discussion}). In this paper, we assume a $\Lambda$CDM cosmology with $h=0.7$, $\Omega_{\Lambda}=0.7$ and $\Omega_m=0.3$ and use $\Zsun=0.02$.

\section{Forming low metallicity stars}\label{sec:lowZ}

The total merger rate $\mathcal{R}$ at merger time $\tm$ is given by
\begin{eqnarray}\label{eqn:NofMgal}
 &&\frac{d\mathcal{R}}{d\log \mgal d\tform d\Zc}= \frac{dN}{d\log \mgal d\Zc} \times \\
&&\int_{t_{\infty}}^{\tm} d \tform \frac{dN_{\mathrm{m}}(t_m-\tform,\Zc)}{dt_f} SFR\left(\tform,\mgal \right) \Psi \left(\tform,\mgal,\Zc\right), \nonumber
\end{eqnarray}
where $\mgal$ is the  galaxy stellar mass, $\Zc$ the progenitor metallicity. We perform the integral over the formation time of the progenitors $\tform$. SFR is the star formation rate  (in $M_{\odot}$ yr$^{-1}$) and $dN/d\log \mgal$ the stellar mass function (SMF) at the time of the merger (per unit comoving volume,  taken from \citealt{Tomczak:2014}). $\Psi$ is the fraction of stellar mass forming at metallicity $\Zc$ with respect to the total stellar mass formed. $dN_{\mathrm{m}}/dt$ is the delay time distribution of massive black hole mergers per unit solar mass. In this section we will determine the distribution of low metallicity stellar mass as a function of host mass and formation time while in \S\ref{sec:mergers} we will determine the number of black hole mergers per unit solar mass $N_m$.

To determine the amount of low metallicity stellar mass in each galaxy, we include stars formed within that galaxy as well as stars that were brought into the galaxy through mergers. While this ``ex-situ'' star formation is around $30\%$ in present day galaxies \citep{Lackner:2012},  the accreted stars are typically formed in lower mass galaxies, which have a lower metallicity. As such, the ex-situ component cannot be neglected for our work.  The study of galaxy merger histories can be done with hydrodynamic cosmological simulations \citep[e.g][]{Maller:2006} or semi-analytic models of galaxy formation within dark matter halos \citep[e.g][]{Cole:2000,Guo:2008,Fakhouri:2008}.  We use the global fit function from \citet{Cole:2008}, based on the Extended Press-Schechter formalism \citep{Lacey:1993}, which provides the redshift dependent mass distribution of progenitor halos that will merge within the main halo by $z=0$. We neglect the evolution of the main halo and galaxy mass between $\zm$ and $z=0$, but for the observed $\zm$ this is extremely small.

For each galaxy mass we consider, we determine the mass of the corresponding dark matter halo using abundance matching by \citet{Behroozi:2013}. Using 10 Myr timesteps, we build up the amount of low metallicity stars within that halo between $z_{\infty}=8$ and $\zm$ according to the merger tree. We neglect stars formed before $z=8$ due to the lack of observational constraints. While the SFR density at such early times is at least two orders of magnitude below its value at the  peak of star formation the low metallicity environment will increase their respective contribution to the total merger rate.
For each timestep, we determine the progenitor halo mass function. For each of the progenitors, we determine the corresponding galaxy mass and SFR at that redshift, again using data from \citet{Behroozi:2013}. We then determine the fraction of stars forming at $\Zc$ using the redshift dependent MZR.  Finally, we add up the low metallicity contributions of all the progenitors to get the total amount of low metallicity stars formed at the considered redshift that will be in $\mgal$ at $\zm$.

We model 11 metallicity bins between $\Zc=0.01\Zsun$ and $\Zc=\Zsun$, each bin being 0.2 dex wide. We specifically examine BBH progenitors formed at $\Zc /\Zsun =$ 0.3, 0.1, and 0.01 ; $\Zc = 0.1 \Zsun$ is broadly the most likely value \citep{Belczynski:2016}, but progenitors form for $\Zc \lesssim 0.5\Zsun$ . The observational determination of gas-phase metallicities, which is needed to tell us where low metallicity stars form, unfortunately, has systematic uncertainties of $\simeq 0.5$~dex  owing to different nebular calibrations \citep{Kewley:2008,Steidel:2014}.

We therefore determine the mean metallicity of the star forming gas using the mass-metallicity relation from \citet{Ma:2016}
\begin{equation}
  \label{eq:MZR}
  12+\log(\mathrm{O/H})=0.35(\log(\mgal)-10)+0.93 \exp^{-0.43z}+7.95.
\end{equation}

This MZR is based on high-resolution cosmological zoom-in simulations suite FIRE \citep{Hopkins:2014_FIRE}, which reproduce the observed stellar mass-halo mass relation, Kennicutt-Schmidt law, star forming main sequence and star formation histories. More importantly, the simulated MZR agrees with both gas phase and stellar metallicity measurements observed at low redshifts for $10^4\leqslant~\mgal\leqslant~10^{11}\msun$ \citep{Tremonti:2004,Lee:2006} as well as the data at higher redshifts \citep{Erb:2006,Mannucci:2009}. This MZR agrees well with the \citet{Pettini:2004} calibration, removing some of the systematic uncertainties. If, however, we systematically increase all metallicities by switching to  the \citet{Kobulnicky:2004} calibration, we obtain the same relative merger rates but lower the total rate by a factor of five.

 To determine the actual amount of low metallicity star forming gas within a galaxy, we need to assess the scatter with respect to the mean metallicity, as increased scatter will increase the number of BBH progenitors. \citet{Tremonti:2004} indicate a scatter with $\sigma\simeq .1$ dex between different galaxies independent of redshift.  This is significantly lower than the scatter derived from Damped Ly$\alpha$ systems (DLA) \citep{Rafelski:2012}. In the latter, galaxy masses are not measured, and their scatter likely accounts for most of the scatter in metallicity \citep{Dvorkin:2015}. A significant scatter may also be present within a given galaxy. In spiral galaxies, the metallicity decreases by about $0.03-0.06$ dex $\mathrm{kpc}^{-1}$ with galactocentric radius \citep{Henry:2012,Berg:2013}. At a given radius, scatter is typically $\simeq 0.1$ dex. Assuming both radial and non-radial variations of $\sigma\simeq 0.2$, we have a total standard deviation of $\sigma=0.3$.  Using a normal distribution for $[O/H]$, we then determine $\Psi$, the fraction of gas at $\Zc$.

\begin{figure}
\centering
\includegraphics[width=\columnwidth]{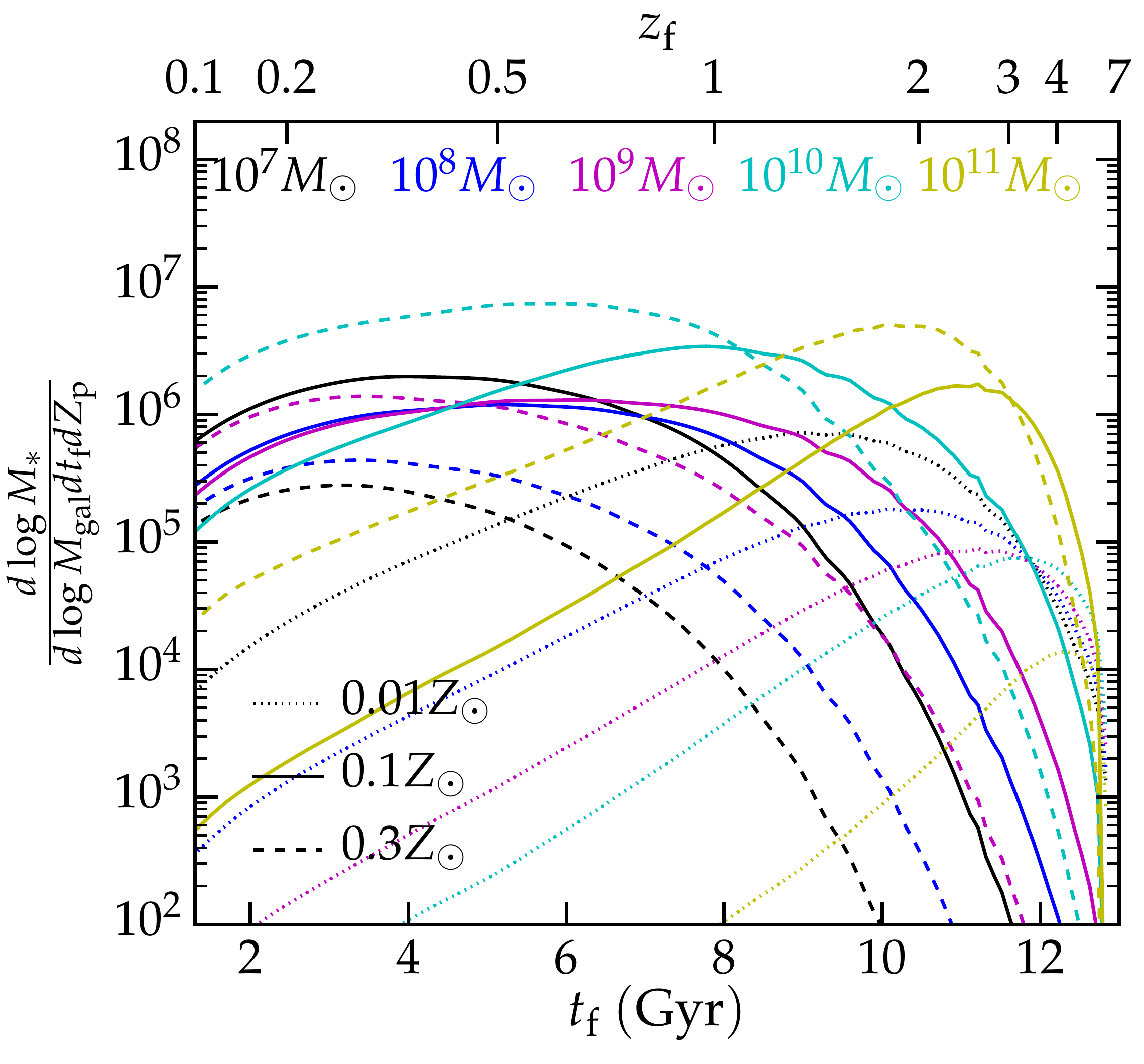}
\caption{Cosmic mass density (in $\msun $ Mpc$^{-3}$ Gyr$^{-1}$ $\msun^{-1}$) of stars at different metallicities (linestyles) in present-day galaxies with a total galaxy stellar mass $\mgal=10^{7-11} \msun$ (color as labeled), as a function of lookback time (redshift) to when the stars actually formed. \label{fig:stars}}
\end{figure}

Fig.~\ref{fig:stars} shows the low metallicity stellar mass density as a function of lookback time (and redshift) to its formation for various galaxy masses (taken at $\zm=0$).  Stars with $\Zc = 0.01 \Zsun$ (dotted lines) form before $z\simeq 2$ and can be found in dwarf galaxies. Stars with $\Zc=0.3\Zsun$ (dashed lines) formed more recently (1$\leqslant \zform \leqslant$ 2) in  Milky Way type galaxies. Stars with $\Zc=0.1\Zsun$ show a combination of both trends. When we neglect galaxy mergers, low metallicity star formation is reduced by at least an order of magnitude and limited to $\zform \geqslant 2$, with little dependence on $\mgal$.  We find that most of the metal poor stars formed at low redshifts were brought in through mergers and  were formed in galaxies smaller that their final host. In the next section we will determine the typical time between progenitor star formation and BBH merger in order to determine from which of these environments GW150914 most likely originated.

\section{Time delays for massive BBH mergers}\label{sec:mergers}
To link the SFR, progenitor metallicity, and host mass evolution discussed above with BBH mergers that are detectable by LIGO, we compute a set of BPS models. Many phases in the evolution of binary stars remain poorly understood and previous BPS studies have shown that this results in large uncertainties in the BBH merger rate \citep[e.g.][]{Lipunov:1997,Sipior:2002,Dominik:2013}. Since this work focuses on host galaxies, and not binary evolution, we consider a simple, single set of standard assumptions consistent with observational constraints. We note that our models do not include the recently proposed massive overcontact binary BBH formation channel \citep{Marchant:2016,Mandel:2016}. We focus on field binaries and neglect BBHs that are dynamically formed in globular clusters \citep[e.g.][]{Downing:2011,Rodriguez:2015}, which would typically form at high redshifts and preferentially reside in more massive galaxies. Lacking observational constraints, we also neglect BBH  stemming from Pop III stars \citep{Kinugawa:2014}, which are not likely candidates for GW 150914 \citep{Hartwig:2016} and which  contribution to  the gravitational wave background is still uncertain \citep{Dvorkin:2016}.

The BPS models are computed with the binary star evolution code \texttt{BSE} described in \citet{Hurley:2002}, which we have updated to improve the treatment of massive binaries. We use the weaker, metallicity dependent wind mass loss prescriptions from \citet{Belczynski:2010}. Updated remnant mass prescriptions are taken from \citet{Belczynski:2008}.  BH birth kicks are modeled following \citet{Dominik:2013}. This results in the production of BBHs with component masses $\ga 25 \msun$ that are not disrupted by powerful natal kicks. The kicks are drawn from a Maxwellian distribution of width 265 km s$^{-1}$, reduced according to the amount of material that falls back after core collapse.

We have also updated the treatment of some mass transfer scenarios in \texttt{BSE}. We force systems that experience a common envelope phase while the mass donor is in the Hertzsprung gap to merge{\footnote{Stars in the Hertzsprung gap lack a steep density gradient between the core and envelope so there is no clear boundary to halt the inspiral of the companion and prevent a stellar merger \citep{Ivanova:2004,Belczynski:2007}.}.  For stars that have evolved beyond the Hertzsprung gap, we take the common envelope efficiency to be unity, and compute the envelope binding energies with the \texttt{BSE}-default, evolutionary-state-dependent formulae. Furthermore, we allow stable Roche lobe overflow mass transfer to be non-conservative and assume that only half of the mass lost by the donor is accreted by the companion \citep{Dominik:2013}. With this updated version of \texttt{BSE} we are able to produce a reasonable estimate for the BBH merger delay time distribution given an initial population of binary stars.

We construct the delay time distribution from a Monte Carlo ensemble of $2.5\times10^6$ binaries. Primary masses range from $25-150 \msun$ and are drawn from the initial mass function (IMF) given by \citet{Kroupa:2001}.  This allows for a wider mass distribution than the GW150914 event, which will be representative for future massive black hole binary detections. When we select a narrow mass range, set by the uncertainties on the GW150914 detection ($M_1=36_{-4}^{+5}\msun$ and $M_2=29_{-4}^{+4}\msun$), we find qualitatively very similar trends. The initial mass ratios and orbital periods are drawn from the distributions measured by \citet{Sana:2012}.  Initial eccentricities are drawn from a thermal distribution $f(e) \propto 2e$. We evolve the same population of binaries  for the 11 metallicity bins we consider.

Fig.~\ref{fig:delaytime} shows the number of BBH mergers per solar mass of stars formed that occur a time $t_\mathrm{delay}$ after the stellar binary forms. We only considered BBH mergers with total mass larger than $40 \msun$. Due to the metallicity dependence of the wind mass loss rates, binaries formed at $\Zc = 0.01 \Zsun$ produce the most massive BHs.  Accordingly, these extremely low metallicity stars have the largest number of massive BBH mergers per unit stellar mass. However, at very late times higher metallicity stars account for a comparable number of mergers.

If we include BBH mergers of all masses (not shown here), $dN_\mathrm{m}/dt$ at each metallicity considered here approaches the standard $t^{-1}$ dependence \citep[e.g.][]{Dominik:2013,Belczynski:2016}. This agreement with previous work is encouraging because, for our purposes, it is most important to properly capture the {\em shape} of the delay time distributions. When we restrict our study to BBH mergers with total mass larger than $40 \msun$, only the $\Zc = 0.01 \Zsun$ delay time distribution $dN_\mathrm{m}/dt$ follows the $t^{-1}$ dependence, as is shown by the flat line for $N_{\mathrm{m}}(t)$.  At higher metallicity, short mergers are absent because of larger stellar radii, which make many systems merge as stellar binaries before producing a BBH. On top of that, some binaries contract less during the common envelope phase, because of the lower envelope binding energy, resulting in BBHs that merge at later times.  Except for the very low metallicity progenitors, we do not expect mergers from recently formed stars.

\begin{figure}
\centering
\includegraphics[width=\columnwidth]{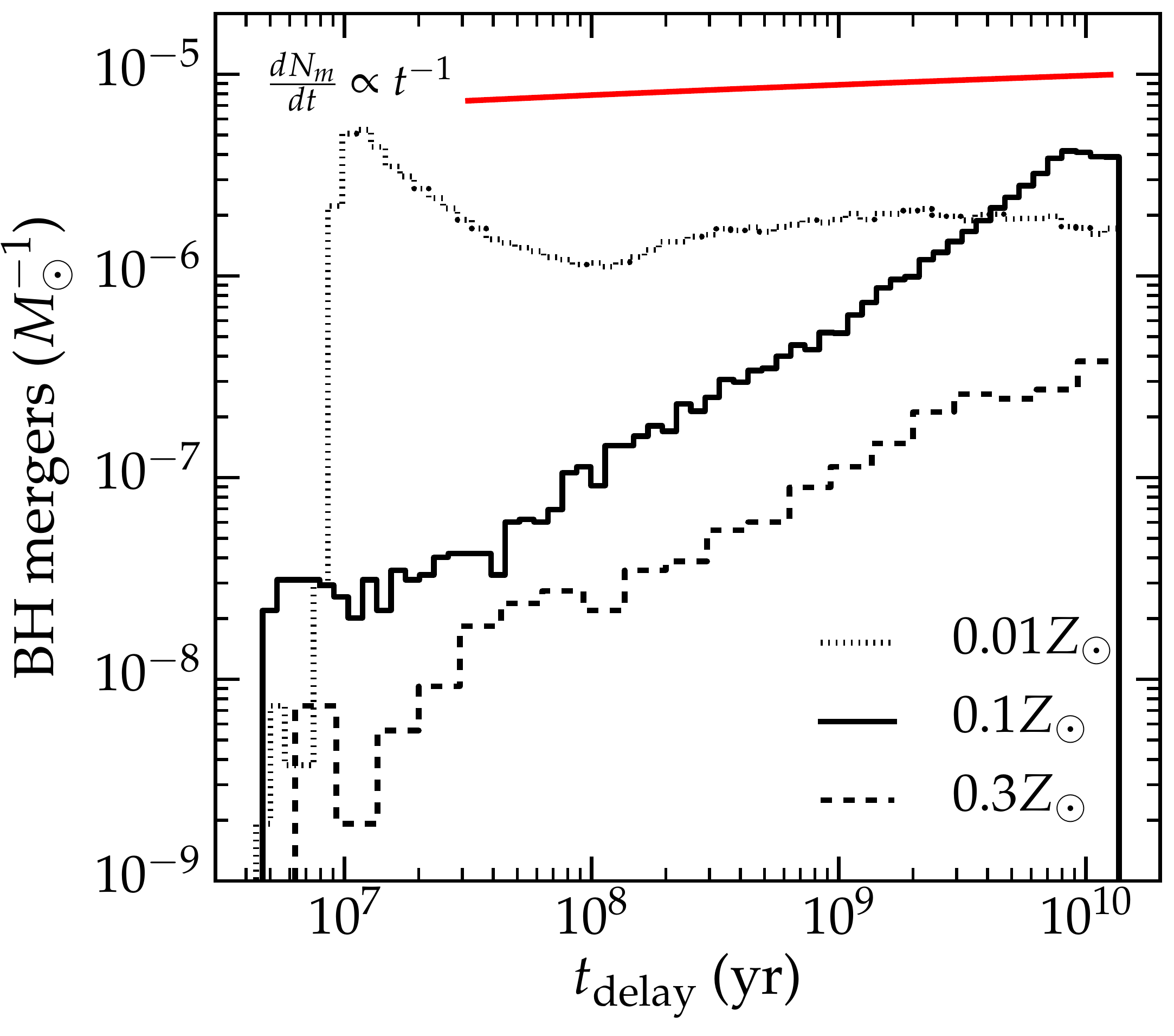}
\caption{Number of massive BBH mergers per solar mass of star formation $N_{\mathrm{m}}$ as a  function of time since formation for a stellar population with a Kroupa IMF and BBH mass $> 40\;M_\odot$. The upper limit in $t_\mathrm{delay}=t_{\mathrm{m}}-\tform$ is the Hubble time. For massive BBH mergers, only the $0.01 \Zsun$ population follows the standard $dN_\mathrm{m}/dt \propto t^{-1}$ evolution, shown with a red line. } \label{fig:delaytime}
\end{figure}

\section{Formation of BBH merger candidates }\label{sec:combine}
We now combine the number of low metallicity stars formed in different galaxies at different epochs with the number of mergers after a certain delay time for different progenitor metallicities (see Eq.~\ref{eqn:NofMgal}).  We assume a binary fraction of $0.7$ \citep{Sana:2012}. Fig.~\ref{fig:mergers} shows the merger rates as a function of host galaxy mass, progenitor formation time and metallicity. The distribution is bimodal with early formation of $\Zc\gtrsim 0.1 \Zsun$ progenitors now present in massive galaxies and lower metallicity progenitors forming later in dwarf galaxies. The latter have limited star formation but are numerous and have a low metallicity. The contribution of dwarf galaxies is sensitive to the extrapolation of the low-mass galaxy SMF below observational completeness but the relatively flat galaxy mass distribution is robust to those uncertainties.

Integrated over all galaxy masses, the formation time of the progenitors is a rather flat distribution over  the last  8 Gyr. We do not recover the strongly bimodal birth time distribution from \citet{Belczynski:2016} because of our more accurate treatment of the star forming gas  metallicity and star formation. Most of the progenitors form around $\Zc\simeq 0.1\Zsun$. Many stars form at higher $\Zc$, but the number of mergers per unit solar mass is drastically reduced.  At lower progenitor metallicity, more systems merge, but the amount of stars formed is low. If we were to include recently proposed fast merger channels \citep{Mandel:2016,Marchant:2016}, the distribution of host galaxies and formation times would be very similar to the distribution of low metallicity stars, with a possible contribution from low redshift galaxies.

\begin{figure*}
\centering
\includegraphics[width=\columnwidth]{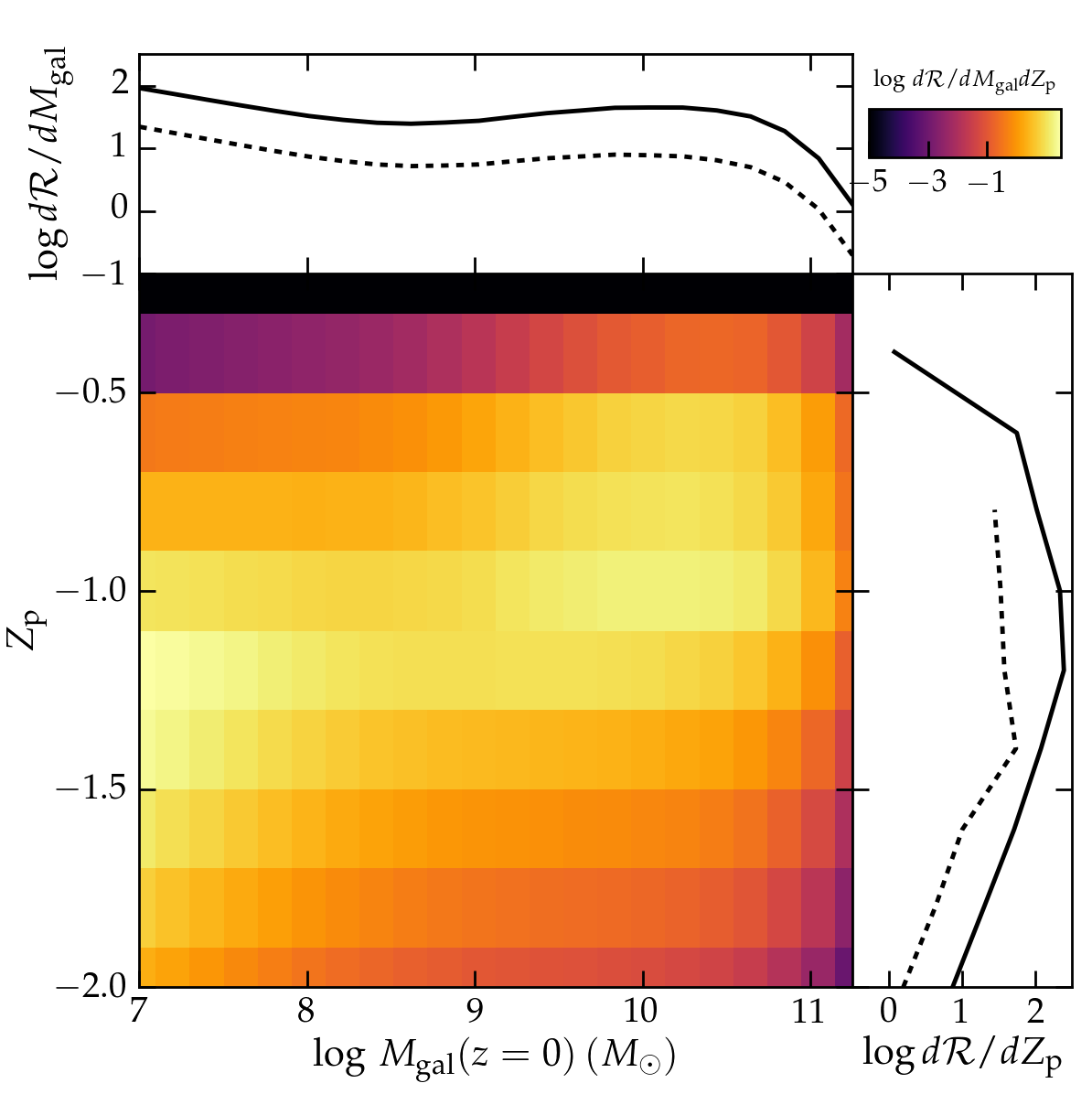}
\includegraphics[width=\columnwidth]{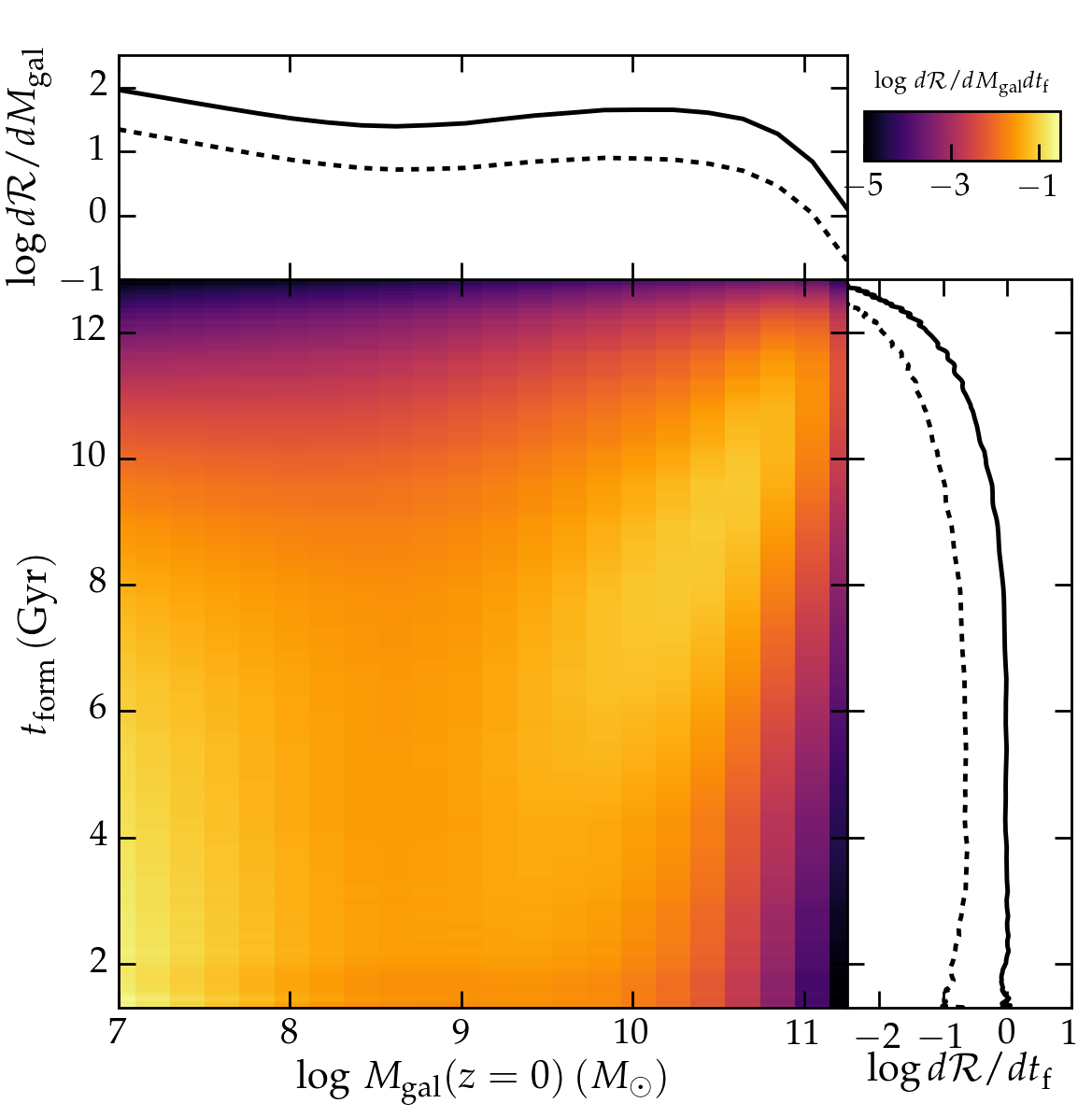}
\caption{Comoving merger rate (Gpc$^{-3}$ yr$^{-1}$) as a function of present day galaxy mass and metallicity (left) and  lookback time to the formation of the progenitor (right). The distribution has  been integrated over formation time (left) and progenitor metallicity (right). Side panels show the integrated rates for our total sample (solid lines) as well as the restricted  GW150614-like  sample (dotted lines).}
\label{fig:mergers}
\end{figure*}

The total merger rate we find is $\mathcal{R}=850$ Gpc$^{-3}$ yr$^{-1}$ for our total mass sample and $\mathcal{R}=150$ Gpc$^{-3}$ yr$^{-1}$ when we restrict ourselves to the exact masses observed in GW150914. After the first observing run, the LIGO estimate of the merger rate of GW150914-like black holes is 2-53 Gpc$^{-3}$ yr$^{-1}$ \citep{LIGO:2016_rate}. While our model is based on standard assumptions for galaxy evolution and massive binary evolution, the total predicted merger rate overestimates the observed rate only by a factor 3. Choosing a metallicity calibration that predicts a lower MZR, a lower binary fraction and/or higher common envelope binding energy will naturally decrease these numbers. As we focus on formation conditions rather than absolute rates, we choose not to fine tune our model.

\section{Discussion and Conclusion}\label{sec:discussion}

In this paper, we compute  when and where  GW150914 most likely formed. Using only the strong constraint on the progenitor's metallicity and combining a state of the art binary population synthesis model with a complete cosmological description for the evolution of low metallicity gas, we find that GW150914 likely formed in a massive galaxy at $1\leqslant\zform\leqslant2$, but later formation in a dwarf galaxy is also possible.  In fact the distribution of BBH merger progenitor formation times is remarkably flat for  $\tform \simeq 1-10$ Gyr, and differs from the strongly bimodal distribution from \citet{Belczynski:2016}.  Their computation is based on the metallicity evolution of DLAs, which ignores the crucial mass dependence of the metallicity.  Our model also includes galactic mergers, which  allow BBH progenitors formed in dwarf galaxies to end up in massive systems at $\tm$ and strongly increase the amount of mergers in the latter.  Still, we find a large contribution of mergers in dwarf galaxies, whic is radically different from the distribution of present day stars, supernovae and BH which are strongly concentrated around $\mgal\simeq 10^{11}\msun$. This work presents the first determination of the formation conditions for the massive BBH mergers we currently observe.  Without fine tuning, the total merger rate we predict is compatible with the LIGO detection rate.

Our work assumes that the only environmental impact on stellar evolution is progenitor  metallicity, allowing us to decouple galactic evolution and stellar evolution, including multiplicity and the initial mass function.  As such, the large uncertainties in massive stellar evolution only affect our absolute merger rate, but not its dependence on galaxy mass and formation time. Unless the metallicity dependence of stellar evolution were to be drastically revised, our model can be easily rescaled for different models of massive stellar evolution.

Uncertainties also affect our model for galaxy evolution, especially in small galaxies at high redshifts where star formation rates and particularly metallicity are very hard to determine observationally.  We have assumed dwarf galaxies form the same amount of massive binaries per unit solar mass than larger galaxies, neglecting the fact that they may not host large enough molecular clouds to do so. As our understanding of high redshift star formation and stellar evolution improves with data from the James Webb Space Telescope (JWST) and the Wide Field Infrared Survey Telescope (WFIRST), our method will become a valuable tool to understand BBH mergers.

As Advanced LIGO and Advanced $Virgo$ reach their design sensitivity \citep{LIGO:2016_detector}, they will detect hundreds of BBH mergers, up to $\zm\lesssim 1$. BBHs merging at these redshifts formed during the peak of cosmic star formation, with a rather flat distribution of galaxy mass.  In this context, this will provide strong tests of our models and the otherwise elusive nature of high redshift star formation and/or the metallicities of high-redshift or faint galaxies. Our method can further be combined with galaxy catalogs to predict typical distance distributions and sky localizations for future detections.

\section*{Acknowledgments}
Support for AL and PFH was provided by an Alfred P. Sloan Research Fellowship, NASA ATP Grant NNX14AH35G, and NSF Collaborative Research Grant \#1411920 and CAREER grant \#1455342. Support for SGK was provided by NASA through Einstein Postdoctoral Fellowship grant \#PF5-160136 awarded by the Chandra X-ray Center, which is operated by the Smithsonian Astrophysical Observatory for NASA under contract NAS8-03060.   DC was supported through the Walter Burke Institute for Theoretical Physics and the Sherman Fairchild Foundation and the Caltech Department of Astronomy. The authors thank Chris Pankow, Evan Kirby, Xiangcheng Ma, Fangzhou Jiang, and  Peter Behroozi for very helpful and stimulating discussions. We thank the referee for a constructive report that improved and clarified the manuscript.

\bibliographystyle{mnras}
\bibliography{LIGO_host}

\end{document}